\documentclass{article}
\usepackage{amsmath}
\usepackage{upgreek}
\usepackage{graphicx}

\begin{document}

\title{Waveguide Resonator with Integrated Phase Modulator for Second Harmonic Generation}

\author{M. Stefszky, \and M. Santandrea, \and  F. vom Bruch, \and  S. Krapick, \and  C. Eigner, \and R. Ricken, \and  V. Quiring, \and  H. Herrmann, \and and C. Silberhorn}

\maketitle





\begin{abstract}
We report second harmonic generation from a titanium indiffused lithium niobate waveguide resonator device whose cavity length is locked to the fundamental pump laser using an on-chip phase modulator. The device remains locked for more than 5 minutes, producing more than 80\% of the initial second harmonic power. The stability of the system is seen to be limited by DC-drift, a known effect in many lithium niobate systems that include deposited electrodes. The presented device explores the suitability of waveguide resonators in this platform for use in larger integrated networks.
\end{abstract}

Practical applications of many optics technologies have been enabled through the use of integrated devices due to the fact that they typically offer improved reproducibility, greater stability, fiber compatibility and stronger nonlinear efficiencies over their bulk counterparts \cite{Kaiser16.O,Luo18.O,Sun12.OL,Sohler08.OPN}. In addition to these benefits, integrated devices also allow for combining multiple functionalities in a single device, thereby reducing complicated multi-part tabletop experiments into single devices. In lithium niobate, for example, one can include passive beamsplitting or polarisation splitting, as well as active switches and a wide range of nonlinear interactions \cite{Sharapova17.NJP,Luo19.SA,Krapick13.NJP}.	

As is often the case, these benefits do not come without a cost. Waveguides typically have higher losses than their bulk counterparts, are more often strongly impacted by production errors, and are sometimes susceptible to photothermal and/or photorefractive effects \cite{Sun17.OE,Santandrea19.NJP,Santandrea20.OE}. Furthermore, these issues are typically exacerbated when using optical fields at higher frequencies, making processes such as high power and/or efficient second harmonic generation (SHG) particularly difficult. Device design, particularly for nonlinear processes, therefore generally requires one to carefully consider all of these effects.

Two  methods for overcoming these issues in waveguide SHG are to use pulsed lasers or to utilise photorefractive resistant waveguides. Pulsed lasers reduce both photothermal and photorefractive effects and also increase the nonlinear interaction strength, thereby allowing one to achieve very high single-pass efficiencies at low average input powers in shorter waveguides \cite{Burghoff06.APL,Zheng02.JOSAB}. Alternatively, one can utilise waveguide systems that do not suffer from photorefractive effects, typically involving some form of doping \cite{Sakai06.OL.SCI,Li13.OE}. These devices enable the use of continuous-wave pump fields at high input powers, which are often preferred over pulsed solutions, for example, in interferometry \cite{GWOQE.NPH}. These photorefractive resistant  devices have been used to demonstrate high single-pass conversion efficiencies \cite{Sakai07.OL,Sun12.OL,Jechow07.OE}. Another advantage to single-pass schemes is that the impact of losses is minimized, due to the fact that the losses are only sampled once.

An alternative solution to achieving high SH conversion efficiencies in waveguides is the use of waveguide resonators. Waveguide resonators allow for high conversion efficiencies of continuous-wave lasers at low input pump powers \cite{Bi12.OE,Luo18.O,Regener88.JOSAB} and provide the ability to tailor the desired operating point of the device through selection of the resonator reflectivities. Additionally, resonator dynamics can be exploited for other purposes, such as the generation of frequency combs, made possible by the intrinsic feedback properties of these systems \cite{Stefszky18Comb.PRA,Zhang19.N}. The drawbacks to resonator systems are that they are highly sensitive to waveguide losses and that they typically require active stabilisation, or locking, of the resonator length for long-term stability.

We recently presented a locked waveguide resonator showing high conversion efficiency and high operating power \cite{Stefszky18.JO}.  However, the stability of the system was optimal when the laser frequency was locked to the resonance condition of the waveguide. Unfortunately many applications, such as large scale networks, will require devices to work at a set wavelength. This could be achieved in the presented system using feedback to the temperature of the waveguide, but the performance of this lock was found to be inadequate.

Due to the presence of the electro-optic effect in lithium niobate, one can deposit electrodes in order to control the phase of the field through application of a voltage. While this is routinely done for single-pass systems and has been investigated in integrated laser systems \cite{Suche95.OL,Amin94.OL}, very little work has been done on integrating phase modulators into waveguide resonator systems for second harmonic generation \cite{Fujimura96.IEEE}. This is possibly due to the known presence of the DC-drift effect often seen in such modulators \cite{Nagata00.IEE} . This effect causes a drift in the operating point (or set-point voltage) of such systems over time and has been the subject of much investigation for single-pass systems \cite{Gee85.APL,Korotky96.JLT,Nakabayashi96.ECOC}. However, the limitations imposed by this DC-drift and the suitability of deposited electro-optic modulators (EOM) for stable resonator operation in lithium niobate systems remains unexplored. 

Here, we present a waveguide resonator device that is locked to the fundamental pump laser frequency using a deposited EOM. The device is locked for over 5 minutes, producing approximately 40$\upmu$W of SH power with a 2mW 1550nm fundamental field entering the cavity. The performance of the device is seen to be limited by the DC-drift effect and the results reveal the requirements to increase the long-term stability of the system.

\section{Device Design and Production}

\begin{figure}
	\centering
	\includegraphics[width=0.8\linewidth]{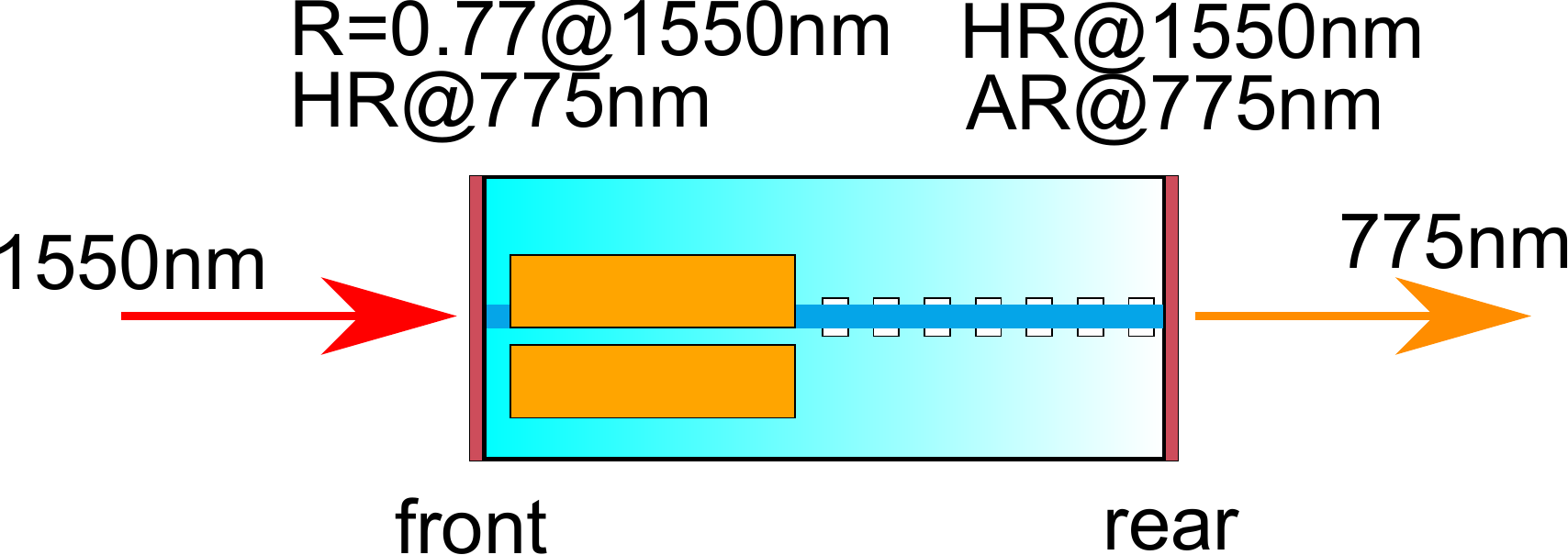}
	\caption{Schematic Layout of the SHG waveguide resonator with integrated phase modulator. The waveguide is represented by the dark blue region, the gold rectangles represent electrodes and the white rectangles represented the poled area. Coatings are deposited on both end facets of the sample. Proportions not to scale.}
	\label{sample}
\end{figure}

A schematic detailing the construction of the device is shown in Fig. \ref{sample}. Standard electric-field periodic-poling with a period of approximately 16.9$\upmu$m is applied to half of the device while the other half contains the deposited electrodes for implementing a phase modulator. Coatings are applied to both facets of the sample in order to define a cavity for the fundamental field and a double-pass configuration for the second harmonic field. This design was selected in order to increase the amount of SH power exiting the desired port of the resonator. The drawback to this design is that the device is interferometrically sensitive to the phase between the forward and reverse propagating SHG waves \cite{Fujimura96.JLT,Imeshev98.OL}.

The waveguides are first produced using a titanium indiffusion method, described in detail in previous work \cite{Stefszky18.JO}. They are defined by depositing titanium stripes onto z-cut LiNbO$_3$ and are designed such that they are single-mode for the fundamental field at around 1550nm.

The next production step is to apply the periodic poling, the standard process of which is described in previous work \cite{Stefszky18.JO}. In order to produce the more complex desired design, a longer sample is first produced, approximately 4cm long, with around 2cm of poling on one end of the device. The poling is done in this way because the liquid electrode periodic poling method used cannot apply poling all the way to the end of the sample. The resulting sample then has approximately 1cm removed from both ends, resulting in a 2cm long sample that is poled along approximately half of its length.

At this point, the waveguide exhibiting the highest performance from the 75 waveguides deposited onto the sample is identified. The losses at 1550nm are measured using a low-finesse Fabry-Perot loss measurement \cite{Regener85.APB}, and the nonlinear performance is determined by measuring both the single-pass SHG efficiency and the profile of the phase-matching function. The profile of the phase-matching function is used as an indicator of waveguide inhomogeneities \cite{Santandrea19.NJP}.

Next, the electrodes are deposited onto the sample. The modulator electrodes consist of a 200nm thick $\text{SiO}_\text{2}$ buffer layer, a 10nm adhesive layer of titanium and finally a 100nm gold layer. The electrodes are deposited by means of confocal DC-magnetron sputter deposition and structured via a standard lift-off process and are deposited onto the section of the sample that has not been periodically poled. Note that, due to the domain inversion resulting from periodic poling, the net effect of a modulator above an (ideal) poled region would be no net phase shift for any applied voltage. 

Finally, end facet coatings are deposited onto the sample using oxygen ion-beam assisted deposition. The mirror coating reflectivities are chosen to provide high SHG efficiency over a range of waveguide losses. This is important because the waveguide losses of the final device are likely to increase slightly from the initial measured values through the many processing steps that are undertaken. The input facet is chosen to have a high reflectivity (HR) coating for the second harmonic wavelength and a reflectivity of $R_{in}=70\%$ for the fundamental field. The output facet is chosen to have an anti-reflection coating for the second harmonic field (R$<1$\%) and a HR coating ($R_{out}>99$\%) for the fundamental field.

\subsection{The Phase Modulator}

For the given crystal orientation (z-cut), phase shifting is achieved by addressing the diagonal tensor element $r_{33}\approx 31$pm/V via a field perpendicular to the sample surface. Consequently, one electrode is situated above the waveguide and the other beside the waveguide, separated by an inter-electrode gap of $G=9\upmu$m and with a width of 400$\upmu$m, as illustrated in Fig. \ref{sample}. 
The half-wave voltage, V$_{\pi}$, that results in a single-pass phase shift of $\pi$ for the TM mode is given by 
\begin{equation}
V_{\pi} = \frac{\lambda G}{n_e^3r_{33} \Gamma L_m}, \label{modeq}
\end{equation} where $n_e$ is the refractive index of the extraordinary axis at the wavelength $\lambda$, $L_m$ is the length of the modulator, and $\Gamma = \frac{G}{V} \int E |A|^2 dA \approx 0.32$ is the overlap integral between the applied electric field $E$ and the optical mode $A$. The overlap integral is calculated using an optical mode profile measured from a similar sample (with an intensity FWHM of 5.3 $\upmu$m and 3.4 $\upmu$m for the horizontal and vertical directions respectively) and the electric field due to the modulator is calculated using \cite{Kim89.JLT}. This expression results in an expected $V_\pi$ of approximately 16V for an 8mm long electrode, assuming that the edge of the electrode is aligned to the edge of the waveguide. Note that the electrode is slightly shorter in length than the desired 1cm due to limitations in processing the end of the sample. The half-wave voltage is equal to the voltage required to scan one free spectral range due to the fact that the field experiences this phase shift twice per round trip. Using this geometry and neglecting the effect of the buffer layer, which would otherwise act to reduce this capacitance \cite{Chung91.JQE}, we estimate the capacitance of the device to be $C\approx6$pF which would give an RC bandwidth of $\Delta f\approx 1.1$GHz, assuming a 50$\Delta$ termination resistance \cite{Tamir}. This bandwidth is much faster than the available electronics and orders of magnitude faster than the bandwidth one requires for a typical cavity lock and as such is of little concern for the desired application. 

\section{Experimental Setup}

\begin{figure}
	\centering
	\includegraphics[width=0.8\linewidth]{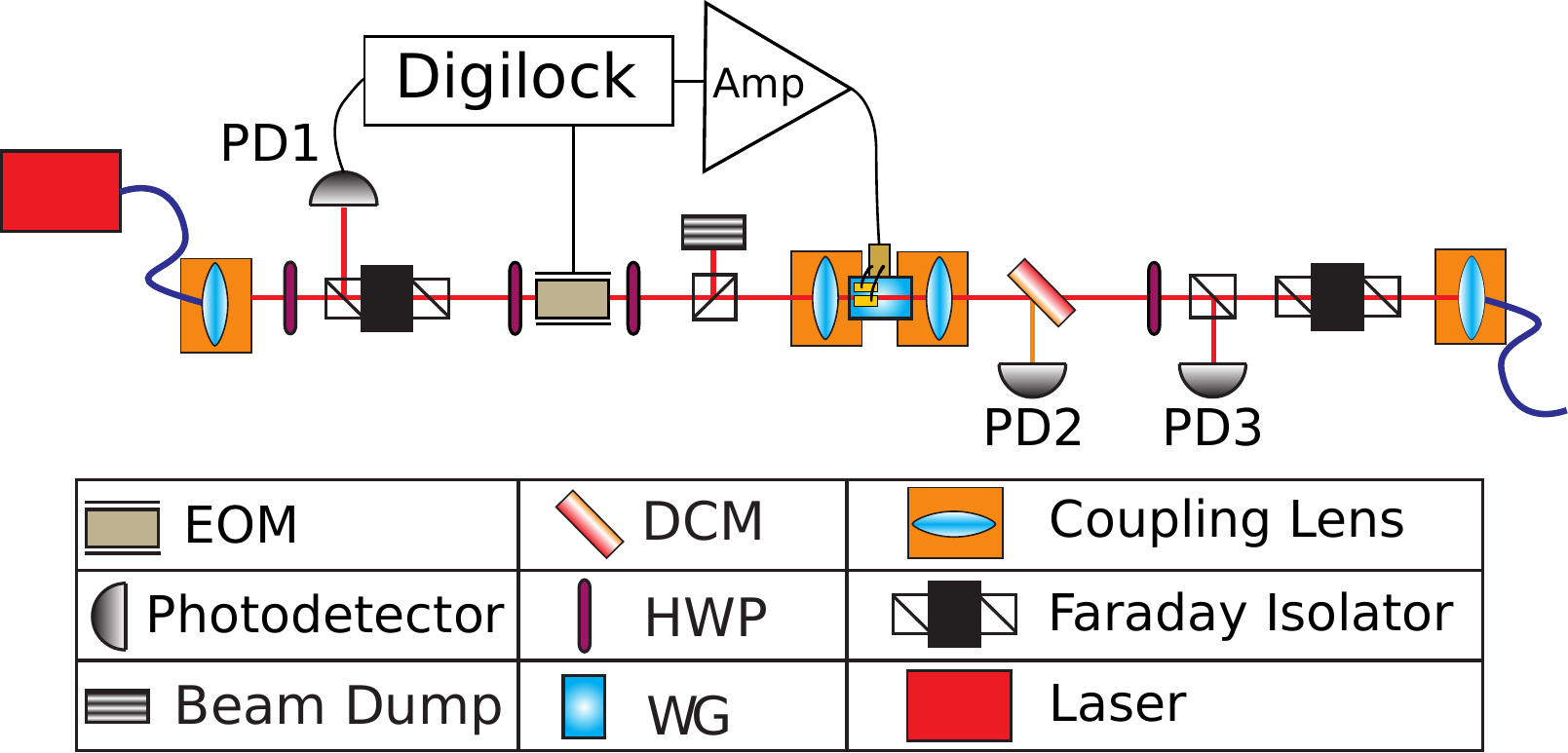}
	\caption{Experimental setup.}
	\label{layout}
\end{figure}

The experimental setup is illustrated in Fig. \ref{layout}. A 1550nm laser continuous-wave (cw) laser (Tunics) is used as the pump source. It passes through a Faraday isolator, an electro-optic phase modulator (EOM) and a polarising beam splitter (PBS) before impinging on the cavity. The electro-optic phase modulator generates sidebands at 25MHz for the Pound-Drever-Hall (PDH) locking scheme \cite{Black01.AJP}. The light that is reflected from the resonator is redirected by the Faraday Isolator and detected on photodiode PD1 to derive the locking signal. The PBS in combination with a half-wave plate provides power control. The waveguide itself is heated to approximately 180$^\circ$ Celsius in order to reduce the effects of photorefraction using a two-stage oven consisting of a resistive heater below a Peltier element which is used for fast fine-tuning of the temperature. After the waveguide a dichroic mirror (DM) separates the generated second harmonic field from the fundamental field. The second harmonic field is directed to photodetector PD2, while the fundamental field can be directed using a half-wave plate PBS combination to either photodetector PD3 or fibre-coupled to a single-mode fibre after transmission through a second Faraday Isolator. The electrodes on the waveguide are connected to a copper strip on the oven and are directed to an amplifier into which the control signal or a triangular sweep signal can be fed.

\section{Linear Characterization}

The linear performance of the device (the losses, coupling efficiency, and mirror reflectivities) at the fundamental wavelength are characterized using the same methods described in \cite{Stefszky18.JO}. Firstly, the reflectivity of the input/front mirror is measured  by translating the sample laterally such that the fundamental field is incident on the substrate. The amount of reflected light is then measured with the aid of the first Faraday isolator and provides an accurate estimate of the reflectivity of the front-facet coating. Next, the reflected and transmitted powers of the fundamental field are measured  when the system is both on and off resonance. These measured values, in combination with the coupling efficiency of the fundamental field into the waveguide, can be used to determine the reflectivity of the output/rear facet and the internal losses of the resonator.

The coupling efficiency of the fundamental field to the waveguide is measured after slight alterations to the experimental setup. Both Faraday isolators are inverted such that light can pass in the opposite direction and the laser is coupled to the fibre at the rear of the waveguide. The resonator is then brought to resonance and the field exiting the waveguide is coupled to the fibre where the laser was initially connected. The amount of light entering this fibre then provides the coupling efficiency of the waveguide mode to this fibre.

Using these characterisation techniques the coupling of the fundamental light into the waveguide was found to be 0.68 $\pm$ 0.03  and the reflectivity of the front facet at this wavelength $R_{in}$=0.694 $\pm$ 0.007. Using these values, it is then found that the losses of the cavity at the fundamental wavelength are 0.25 $\pm 0.01$ dB/cm and the reflectivity of the rear facet is $R_{out}$=0.997 $\pm$ 0.001. We note that initial characterization of the waveguide using the low-finesse Fabry-Perot method estimated the losses at around 0.16dB/cm, much higher than what has been previously demonstrated \cite{Luo15.NJP} due to the need to find a waveguide with both low losses and desirable nonlinear properties. The loss measurements reveal that the losses have increased due to the presence of the electrode. Minimization of these additional losses by varying the properties of the modulator is currently under investigation. Given the length of the sample of $L=$2cm, the  free spectral range $FSR = \frac{c}{2 L n_{g}} = 3.47$GHz, finesse $\mathit{F} = \frac{\pi \sqrt{r}}{1-r}=10.4$ and cavity linewidth $\nu = \frac{FSR}{\mathit{F}} = 334$\,MHz can all be determined; where $n_g$ is the group index of the fundamental wave, $c$ is the speed of light, and $r = \sqrt{R_{in} R_{out} e^{- \alpha_\omega L}}$ with fundamental field (intensity) losses of $\alpha_\omega$. Note that the described cavity provides an intra-cavity power enhancement of the fundamental field given by $f=\frac{1-R_{in}}{\left( 1-\sqrt{R_{in}R_{out}} e^{a_{\omega}L} \right)} \approx 20$, significantly enhancing the system efficiency in comparison to a single-pass device \cite{Regener88.JOSAB}.

\section{Nonlinear Characterization}

The nonlinear performance of the device is then characterized. The expected nonlinear conversion efficiency $\eta$, defined as the ratio of the output second harmonic power $P_{SH}$ to the input fundamental field power $P_{in}$, is calculated using an extension to the theory presented by Berger \textit{et al}\cite{Berger97.JOSAB}. An overview of the treatment is presented here, while a more detailed treatment can be found in a paper dedicated to investigation of this extended theory \cite{Santandrea20.ArX}. One first finds the circulating fundamental field traveling in the forward direction at the input mirror $E^f_\omega(0)$  using the standard Fabry-Perot treatment
\begin{eqnarray}
E^f_\omega(0) &=& E_{in} \frac{\tau_{\omega,0}}{1-\rho_{\omega,0}\rho_{\omega,L}\cdot e^{-i 2 k_\omega L} \cdot e^{-\alpha_\omega L}}
\end{eqnarray}
where we define the wavevector of the fundamental field $k_\omega [m^{-1}]$, the fundamental field (intensity) losses $\alpha_\omega [m^{-1}]$, the complex reflectivity for the input/output facets $\rho_{\omega,0/L}$ at $\omega$, the complex transmission of the input facet $\tau_{\omega,0}$ at $\omega$, and the driving field at $\omega$, $E_{in}$. Assuming negligible fundamental-field pump-depletion over a single round-trip, one can describe the generated second harmonic field amplitude over a single pass of the sample with length $z$ using
\begin{eqnarray}
\frac{d E_{2\omega}}{d z} &=& i \gamma \left[E_{\omega} e^{\alpha_\omega z/2}  \right]^2 e^{i \Delta k z}-\frac{\alpha_{2\omega}}{2} E_{2\omega},
\end{eqnarray}
where $E_{2\omega}$ describes the second harmonic field envelope, $\alpha_{2 \omega}[\text{m}^{-1}]$ is the intensity losses of the second harmonic field, $\gamma [\text{m/V}]$ denotes the nonlinear coupling strength, and $\Delta k = 2 k_\omega-k_{2\omega} +k_{QPM}[\text{m}^{-1}]$ represents the wave vector mismatch between the two fields, where $k_{2\omega}$ and $k_{QPM}=\frac{2 \pi}{\Lambda}$ represent the wave vectors of the second harmonic field and any periodic poling $\Lambda$ respectively. Under the necessary constraints the nonlinear coupling strength can be related to the commonly used single-pass, lossless, normalized efficiency $\eta_0 [1/(\text{W cm}^2)]$ over a length $L$ given by $\eta = \text{tanh}^2\left[ \sqrt{\eta_0 P_{in}}L\right]$ \cite{Regener88.JOSAB} using $\gamma = \sqrt{\eta_0\cdot 10^4}$. With the field evolution described, one can then find a set of equations that describe the second harmonic field at all interfaces within the sample in order to find a self-consistent solution for the second harmonic cavity field.

The SHG conversion efficiency is then measured by monitoring the amount of generated SH power as the fundamental field power is varied. The frequency of this laser is scanned by applying a $\pm$20V triangle waveform to the piezoelectric transducer of the external cavity of the laser which shifts the frequency by approximately -300MHz/V. This applied voltage is sufficient to scan over one free spectral range of the system and removes the need to lock the system with different input powers for each measurement. The SH conversion efficiency is then found by scanning the temperature of the waveguide at each new power to find the maximum SH output power, measured on PD2, for each given input power.

\begin{figure}[ht]
	\centering
	\includegraphics[width=0.6\linewidth]{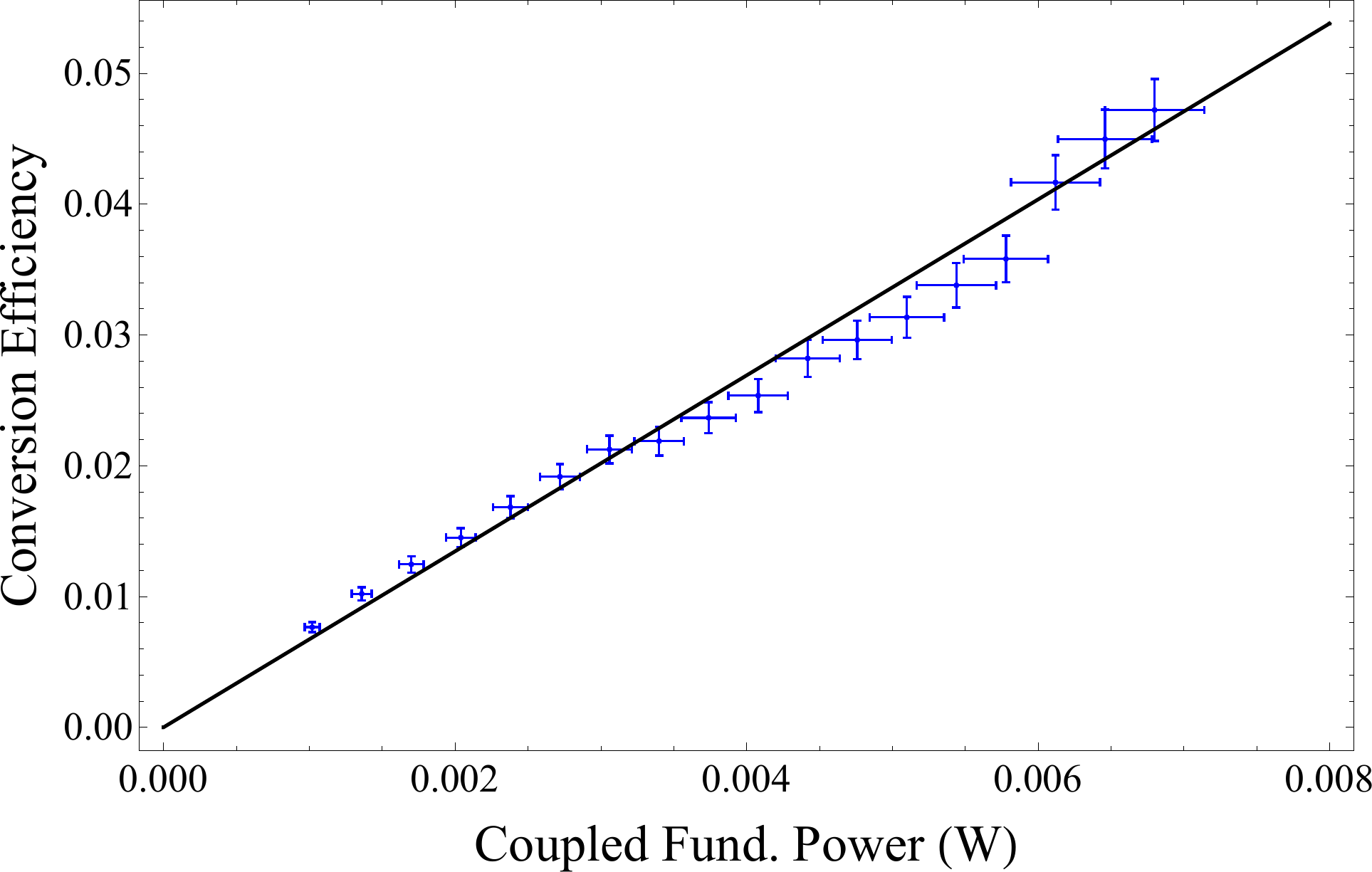}
	\caption{SH Power exiting the waveguide resonator as the fundamental power entering the resonator is varied. The black line shows the predicted SHG with a nonlinear efficiency of 11\% /(W$\cdot$cm$)^2$ using the presented theory, while the blue lines indicates the measured SH powers with associated error bars.}
	\label{SHGMeas}
\end{figure}

The measured conversion efficiency is compared to theory in Fig.\ref{SHGMeas}. The theory is seen to closely predict the conversion efficiency of the device when the only free parameter, the nonlinear efficiency, is set to $\eta_0 = 11\%/(\text{W}\cdot \text{cm}^2)$. This value is much lower than the theoretical maximum of approximately $40\%/(\text{W}\cdot \text{cm}^2)$ for short samples that has been observed in previous samples \cite{Stefszky18.JO}. Note that this theoretical maximum reduces as the length of a sample is increased due to accumulative effects from waveguide inhomogeneities \cite{Santandrea19.NJP}. The observed reduction in the nonlinear efficiency is likely due to a degraded poling quality near the edge of the poled region in the middle of the sample due to field edge effects.

\section{Modulator Performance}

\subsection{Scanning the Phase}

\begin{figure}[ht]
	\centering
	\includegraphics[width=0.6\linewidth]{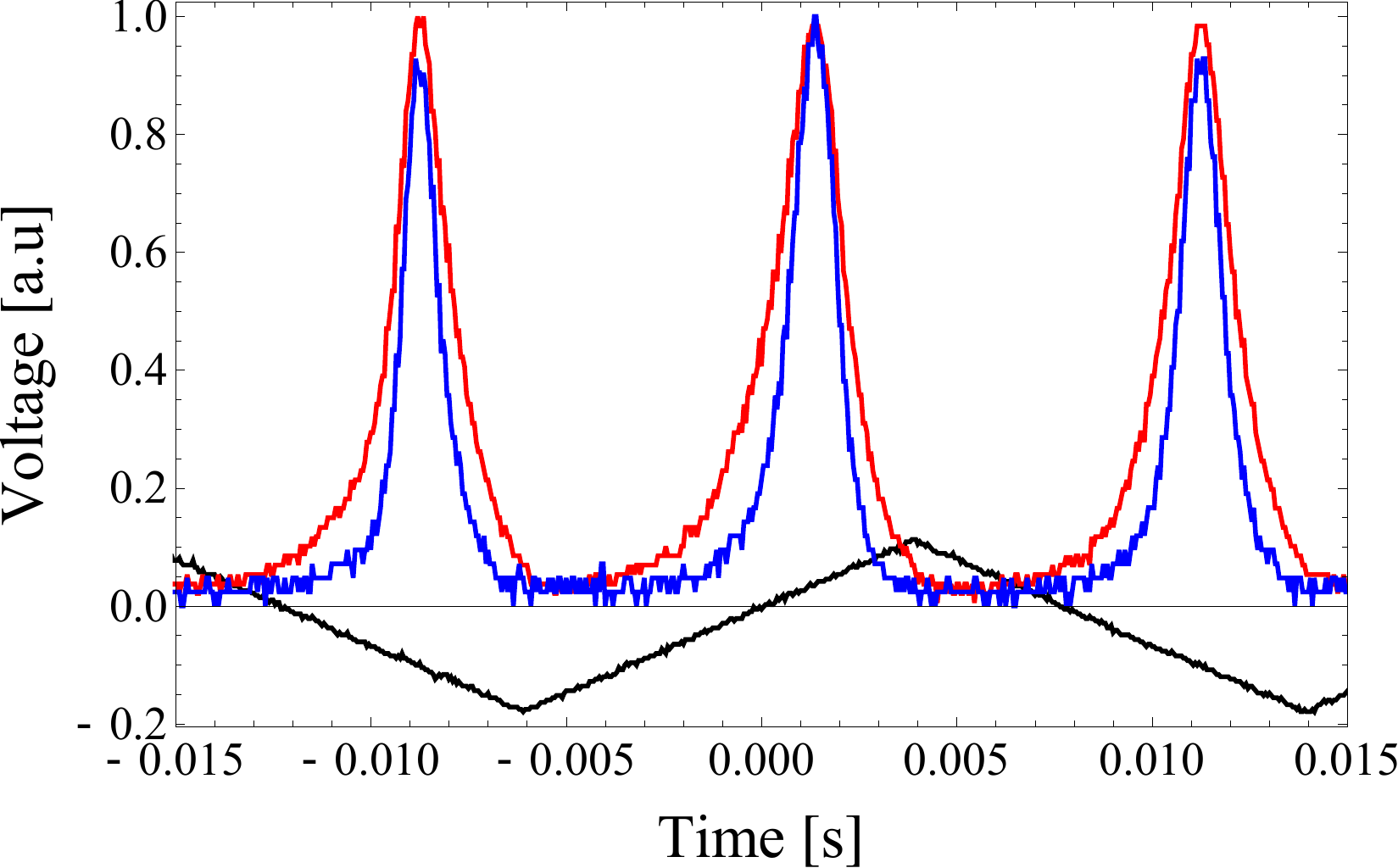}
	\caption{Transmitted fundamental (red) and second harmonic (blue) fields detected on photodetectors PD3 and PD2, respectively, as the swept voltage applied to the modulator is monitored (black line, applied voltage/100). }
	\label{ModScan}
\end{figure}

The performance of the modulator is first investigated by applying a voltage and monitoring the transmitted SH and fundamental fields as approximately 2mW of fundamental power enters the cavity. As the voltage is varied and the resonance condition in scanned, we expect to see the transmitted fundamental power follow the standard cavity response and the second harmonic power follow the square of the amplitude of this field. The resulting transmitted fundamental and SH fields, measured on PD3 and PD2 respectively, are shown in Fig. \ref{ModScan}.

It can be seen that the modulator is able to tune the resonance condition of the cavity over more than one linewidth. It is not possible to directly measure the voltage required to scan one FSR of the system due to the possibility of arcing between the electrodes, but it can be inferred by relating the FWHM and finesse of the cavity. From Fig. \ref{ModScan}, the voltage required to scan over the FWHM of the fundamental resonance condition is found to be approximately 4.3V. Multiplying this by the measured finesse of the fundamental cavity reveals that approximately 45V is required to scan one FSR of the system.

We note that this inferred voltage required to scan one FSR is significantly higher than the expected value and that the transmitted fields show some asymmetry in the profile over the scan. Monitoring the transmitted fields when instead scanning the frequency of the laser with identical fundamental field power reveals that these distortions are no longer present. These asymmetries are therefore attributed to a nonlinearity in the phase shift generated by the EOM, and not to a photothermal or photorefractive effect. The unexpectedly high voltage required to scan one FSR also indicates the presence of unwanted effects. These effects are explored in further detail in the following sections.

\subsection{Locking the Phase}

The length of the cavity is locked to the resonance of the fundamental field using a standard Pound-Drever-Hall locking scheme. This field is modulated at 25MHz using a free-space electro-optic phase modulator and the field is detected upon reflection from the cavity at PD1 to generate the feedback signal (see Fig. \ref{layout}).  The feedback signal is inserted into a Toptica Digilock digital locking box where the feedback signal is tailored using a standard PID configuration. The generated control signal is fed into an amplifier that then drives the phase modulator electrodes that have been deposited on the chip.

\begin{figure}
	\centering
	\includegraphics[width=0.6\linewidth]{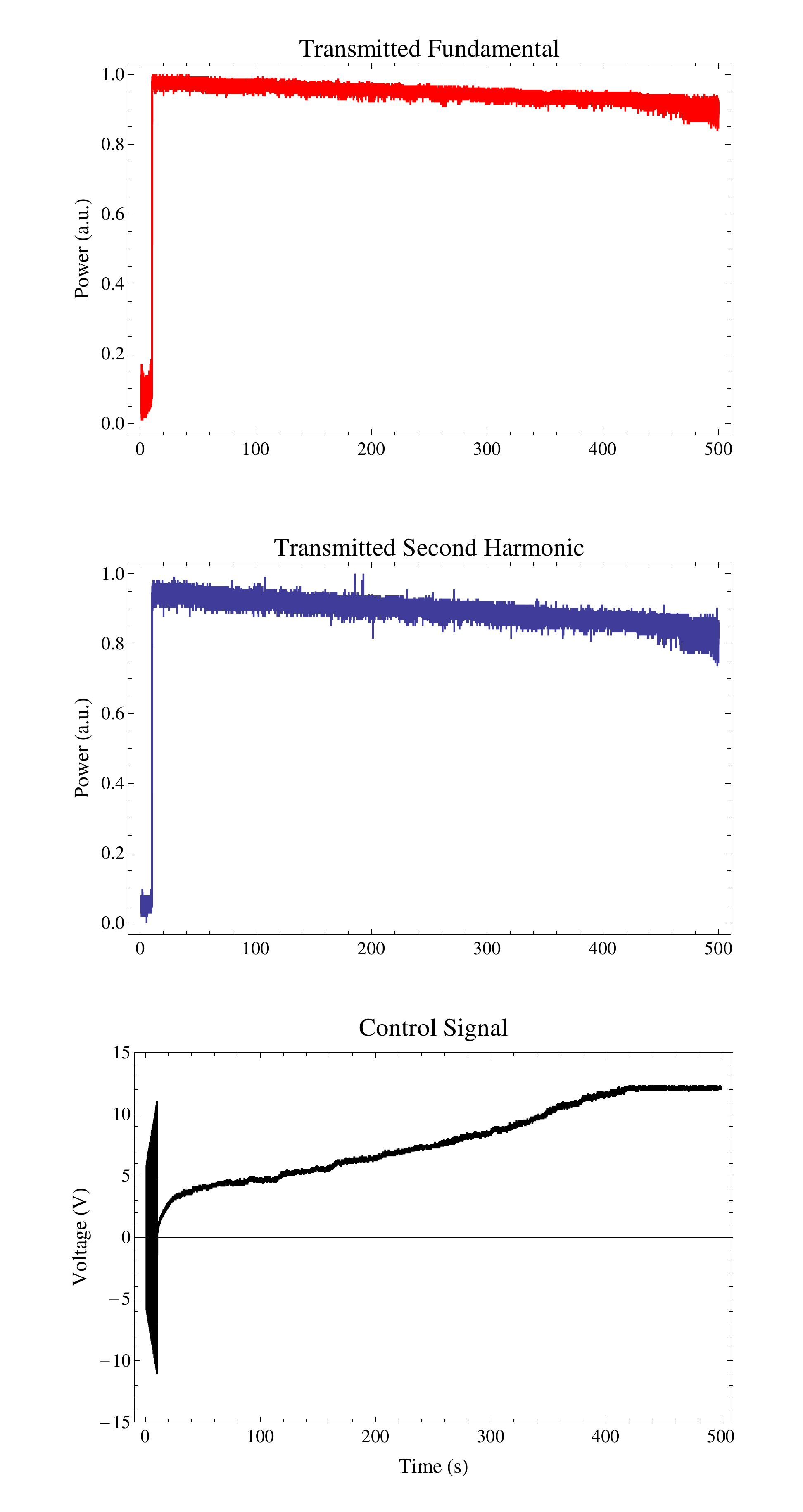}
	\caption{The transmitted fundamental and second harmonic fields (measured on PD3 and PD2 respectively) and the associated control signal from a lock with approximately 2.8mW of fundamental field entering the cavity. The fields are initially blocked and the lock is engaged at approximately 20 seconds.}
	\label{locking}
\end{figure}

The resulting transmitted fundamental and second harmonic fields and the control signal that is fed into the modulator for a typical lock are illustrated in Fig. \ref{locking}. For the first $\approx$ 20 seconds the input fundamental field is blocked and the EOM is driven with a ramp signal. After this time the fundamental field is opened and the system is locked. One can see that there is an accumulating shift in the system operating point, as the control signal increases over time. At $\approx$ 420 seconds the maximum range of the lock (12V) is reached (chosen to ensure that arcing does not occur) and the lock drops. One can see that the cavity system is extremely stable as the cavity remains near to resonance even after the lock is no longer able to compensate for the drift. The slight decrease in the amount of transmitted fundamental power over the measurement was attributed to a reduction in the amount of power transmitted through the waveguide, observed immediately after the measurement time, and not a variation in the locking point of the system. This reduction in the transmitted power could not be corrected for by reoptimising the input coupling and therefore may indicate a slight increase in the internal losses of the system over time, the source of which could not be identified.

The system remains locked for over 5 minutes producing over 80\% of the initial SH power of $\approx 40 \upmu$W. These results show that this platform is already capable of mid-term stability but also reveals that there is room for further improvements.  It is evident that there are currently two limitations to the locking performance of the system; the system drift that can be observed in the control signal, and the limited range of the lock.

\section{EOM DC-Drift}

The control signal shown in Fig. \ref{locking} reveals that two effects with different time constants dominate the dynamics of the system drift. Although the quantitative behavior of the control signal can vary from lock to lock, the general shape of the lock remained mostly unchanged for almost all locks. There is a rapid response to the lock that stabilizes rather rapidly (here in $\approx$ 10-20 seconds) and then a slower response that slowly shifts the operating point of the system until the maximum driving voltage is reached and the lock is unable to further compensate for this shift (here at $\approx$ 400 seconds).

To investigate the source of these drifts the fundamental field power entering the cavity was reduced by nearly an order of magnitude (to $\approx 0.35$mW) and the lock was engaged. The magnitude of the initial rapid response of the system is noticeably reduced but the slower drift component is mostly unaltered, with the lock reaching its range after approximately 500 seconds. Furthermore, the drift is seen to not correlate with a change in the temperature of the sample. Both photorefractive and photothermal effects are power dependent \cite{Sun17.OE} and so are unlikely to be causing this long-term drift. The lock is also unlikely due to a laser frequency drift as the direction of the observed drift and its strength is noticeably correlated to the polarity applied to the EOM when the lock is engaged, with the lock reaching the range limit at -12V in approximately 200 seconds. It is therefore likely that this drift is due to the DC-drift seen in similar systems, originating from charge migration due to the EOM design, particularly the buffer layer choice \cite{Nagata00.IEE,Salvestrini11.JLT}. There are known techniques for reducing this effect, such as varying the chosen buffer layer \cite{Nagata00.IEE,Gee85.APL}, which need to be explored in future iterations of the device.

\section{Locking Range}

The second issue observed in the system is that the voltage required to scan one FSR is approximately 45V, while the expected value is around 16V, as determined through Eq. \ref{modeq}. To investigate this issue further the transmitted fundamental and second harmonic fields were monitored as the applied voltage ramp was varied, as in Fig \ref{ModScan}. A detrimental effect was seen when the applied voltage ramp did not include a DC offset (the polarity of which was insignificant). In the absence of a DC offset the range of the scan was seen to reduce over a few tens of seconds, even though the applied voltage did not vary. At this time the FWHM was seen to correspond to a voltage scan of around 20V, in contrast to the 4.3V seen in the presence of a DC offset. It therefore appears as though there is a build up of slow-moving charge that act to screen the applied voltage in the absence of this DC offset. With the DC offset it appears as though these charges are able to dissipate via some mechanism, thereby removing or reducing the screening effect. This effect warrants further investigation, and may be the cause of the observed high voltage required to scan one FSR, but is outside the scope of this paper.

\section{Conclusion}

In conclusion we have presented a titanium indiffused lithium niobate resonator device that includes an electro-optic modulator for cavity length stabilisation. The device is locked for over 5 minutes and the produced SH power remains within over 80 \% of the initial value. The limitation to the lock is the drift of the resonance condition that occurs over several minutes and appears to be due to the DC-drift effect, which can be minimised in future iterations through optimisation of the modulator design. These pioneering results demonstrate that integrated EOM's can be used to lock the resonance condition of titanium indiffused waveguide resonators, even in the presence of photorefraction, thereby establishing the prospect of using this technology in applications that require operation at a fixed wavelength, such as in large integrated networks. 









\end{document}